\documentclass[preprint,1p,11pt]{ISAS_IR}
\usepackage{abbreviations}

\usepackage{lineno,hyperref}
\usepackage[utf8]{inputenc}
\usepackage{graphics}
\usepackage{epsfig} 
\usepackage[keeplastbox]{flushend}
\usepackage{amsmath}
\usepackage{amssymb}
\usepackage{bm}
\usepackage{textcomp}
\usepackage[linesnumbered,ruled,vlined]{algorithm2e}
\usepackage{url}
\usepackage{color}
\usepackage[labelfont=bf]{caption}
\usepackage{graphicx}
\usepackage{theorem}
\usepackage{booktabs}
\usepackage[english]{babel}
\usepackage{dsfont}
\usepackage{mathtools}
\usepackage{subcaption}
\usepackage{adjustbox}
\usepackage[cal=pxtx,scr=boondox]{mathalpha}
\theoremheaderfont{\bfseries}
\theoremstyle{plain}
{\theorembodyfont{\itshape} }
{\theorembodyfont{\itshape} }
{\theorembodyfont{\itshape} \newtheorem{Example}      {Example}   [section]}
{\theorembodyfont{\itshape}}
{\theorembodyfont{\itshape} }
{\theorembodyfont{\itshape} \newtheorem{Definition}   {Definition}[section]}

\newcommand{\CvM}{Cramér--von Mises\xspace}
\newcommand{\vMF}{von Mises--Fisher\xspace}
\newcommand{\rar}{R\&R\xspace}

\newcommand{\rux}{\mathring{\ux}}
\newcommand{\rX}{{\mathring{\mathbb{X}}}}

\newcommand{\nts}{{n_{\circ}}}

\newcommand{\rmF}{{\mathring{\mF}}}

\newcommand{\cl}{{\mathscr{l}}}
\newcommand{\rom}{\mathring{\omega}}

\newcommand{\mH}{\mathscr{H}}

\newcommand{\neqref}[1]{\,\eqref{#1}}
\newcommand{\eqwith}{{\quad\text{with}\quad}}
\newcommand{\eqand}{{\quad\text{and}\quad}}

\begin{document}
	
	\begin{frontmatter}

	\title{Hyperspherical Dirac Mixture Reapproximation}
	
	\author{{Kailai~Li}, {Florian~Pfaff}, and {Uwe~D.~Hanebeck}}
	
	\address{Intelligent Sensor-Actuator-Systems Laboratory (ISAS)\\
	Institute for Anthropomatics and Robotics\\
	Karlsruhe Institute of Technology (KIT), Germany\\
	emails: kailai.li@kit.edu, florian.pfaff@kit.edu, uwe.hanebeck@kit.edu}

	\begin{abstract}
    We propose a novel scheme for efficient Dirac mixture modeling of distributions on unit hyperspheres. A so-called hyperspherical localized cumulative distribution (HLCD) is introduced as a local and smooth characterization of the underlying continuous density in hyperspherical domains. Based on HLCD, a manifold-adapted modification of the Cramér--von Mises distance (HCvMD) is established to measure the statistical divergence between two Dirac mixtures of arbitrary dimensions. Given a (source) Dirac mixture with many components representing an unknown hyperspherical distribution, a (target) Dirac mixture with fewer components is obtained via matching the source in the sense of least HCvMD. As the number of target Dirac components is configurable, the underlying distributions is represented in a more efficient and informative way. Based upon this hyperspherical Dirac mixture reapproximation (HDMR), we derive a density estimation method and a recursive filter. For density estimation, a maximum likelihood method is provided to reconstruct the underlying continuous distribution in the form of a von Mises--Fisher mixture. For recursive filtering, we introduce the hyperspherical reapproximation discrete filter (HRDF) for nonlinear hyperspherical estimation of dynamic systems under unknown system noise of arbitrary form. Simulations show that the HRDF delivers superior tracking performance over filters using sequential Monte Carlo and parametric modeling.
	\end{abstract}
\end{frontmatter}

\section{Introduction} \label{sec:intro}
On-manifold control and parameter estimation problems have gained growing research interest in recent years~\cite{silvere2013stochastic,li2014unified,casau2020hybrid,de2021dual}. Random variables on unit hyperspheres\footnote{For conciseness, we also denote the domain $\Sbb^2\subset\R^3$, i.e., the unit sphere, as unit hypersphere in this paper.} $\Sbb^{d-1}\subset\R^{d}$ ($d\geq3$) are ubiquitous, e.g., unit vectors denoting spatial orientations~\cite{straub2017manhattan}, unit quaternions parameterizing rigid body rotations~\cite{ECC19_Li}, or vector space models representing text documents~\cite{Wilson2014,banerjee2005clustering}. Probabilistic modeling and state estimation on hyperspheres plays a fundamental role for various control-related application scenarios, particularly robotic perception and navigation, attitude stabilization, and system identification, among others~\cite{feiten2017mpg,tsiotras2019dual,Chaturvedi2011,Hashim2019}.

Mathematically speaking, the unit hypersphere is a compact Riemannian manifold with constant curvature~\cite{hauberg2018directional}, thereby confining the dispersion of random variables to the  hyperspherical geometry. Consequently, probabilistic modeling and inference on hyperspheres are nontrivial due to the underlying nonlinear topology. Existing solutions often rely on local linearization via logarithm/exponential maps to establish a tangent plane, on which the on-manifold uncertainty in the vicinity can be modeled by a Gaussian distribution~\cite{hauberg2013unscented,forster2016manifold,bloesch2017rovio}. Based thereon, classic filtering approaches, e.g., the Kalman filter and its derivatives~\cite{julier2004unscented}, can be adapted~\cite{brossard2020Code} in a conventional manner. Theoretically, an on-tangent-plane Gaussian distribution is warped on the unit hypersphere after undergoing the exponential map. To alleviate this undesirable effect in practice, an assumption of small perturbations is required. Under large uncertainties or fast state transitions, however, artifacts of the warped on-manifold probabilistic interpretation emerge, yielding poor tracking accuracy and robustness for nonlinear hyperspherical estimation~\cite{IFAC20_Li-UPF,LCSS21_Li}.

To model the hyperspherical uncertainties in a topology-aware fashion, distributions from directional statistics have attracted attention recently~\cite{mardia2009directional,JSS19_Kurz,feiten20096d,feiten2017mpg}. Popular ones for hyperspherical filtering include the \vMF distribution with an isotropic dispersion, the Watson distribution with a dispersion of isotropy and antipodal symmetry, and the Bingham distribution, which is non-isotropic and antipodally symmetric. Defined on the unit hypersphere, these distributions inherently obey the geometric structure of the underlying manifold, thereby avoiding local perturbations.

Probability density functions (PDFs) from directional statistics provide a theoretically sound solution to model uncertainties on hyperspheres. However, several issues arise when deploying them to nonlinear hyperspherical estimation in practice. First, almost all of them have no analytic solution to compute the normalization constant. To alleviate computation-intensive numerical approximations, some accelerations were introduced, e.g., exploiting precomputed lookup tables or saddle points concerning the Bingham distribution~\cite{glover2012monte,MFI14_Gilitschenski}. However, insufficient accuracy or numerical instabilities can still occur. Further, stochastic filters based on directional statistics typically require the assistance of samples~\cite{Glover2014QBF}. To cope with strong nonlinearities or high-dimensional states, a large number of samples is often desirable for filtering. While random samples lack efficiency and cannot guarantee reproducible results, deterministic sampling methods are preferable. For instance, deterministic sampling methods in the sense of unscented transform have been developed for \vMF-based hyperspherical~\cite{SPL16_Kurz,MFI20_Li,Sensors21_Li} and Bingham-based quaternion estimation~\cite{TAC16_Gilitschenski,ECC19_Li,Fusion19_Li}. Here, samples are drawn by preserving the first- and second-order moments of the underlying distribution and approximating higher-order shape information if sample sizes are configurable. Though improved performance for hyperspherical estimation was shown, the sampling schemes often involve an optimization or a numerical solver, which increases the computational cost.

More importantly, distributions of parametric form impose strong assumptions on the nature of uncertainties. Empirical data collected from stochastic processes are often of large size. Raw samples can be directly deployed to density propagation in sequential Monte Carlo methods~\cite{arulampalam2002tutorial} for non-parametric estimation. However, a large number of random samples are required for an adequate representation of the underlying uncertainty, and propagating all of them through system dynamics is inefficient. Since unit hyperspheres are compact and bounded, grid-based discrete representations were proposed for non-parametric probabilistic modeling and filtering~\cite{ECC20_Li,IFAC20_Pfaff,Fusion20_Pfaff}. However, the layout of hyperspherical grids is non-adaptive to the underlying geometry, thereby lacking efficiency for modeling uncertainties of complex shape.

Dirac mixtures provide an intuitive solution for discrete modeling of arbitrary continuous distributions based on locations and the associated weights of each component. However, Dirac mixture approximations using raw sample sets from empirical data contain randomness and are inefficient for probabilistic modeling. Therefore, models with a smaller size of components at more representative locations are preferable for non-parametric modeling.

In~\cite{AT15_Hanebeck}, a Dirac mixture reduction scheme was proposed. The \CvM distance (CvMD) was modified to measure the statistical divergence between two multivariate Dirac mixtures based on the localized cumulative distribution (LCD). A reduced sample set is then obtained by optimally preserving the probability mass in the sense of least CvMD. The approach involves approximation for computing the proposed distance measure and relies on selecting a fixed weighting scheme. Also, it requires that the means of the given and the approximated samples are equal. Some extensions to certain directional domains were done, e.g., on the $2$-sphere~\cite{IFAC20_Frisch} or the dual quaternion manifold parameterizing planar motions~\cite{Fusion20_Li}. Both of the extensions are domain-specific. In particular, the approach for $\Sbb^2$ is rather a heuristic modification of its linear original~\cite{AT15_Hanebeck}, thereby suffering from similar issues as mentioned before. Also, the solution relies on spherical coordinates that are difficult to generalize to higher dimensions.

In this paper, we establish a generic and theoretically sound scheme, the hyperspherical Dirac mixture reapproximation (HDMR), for efficient discrete modeling of unknown distributions on $\Sbb^{d-1}\subset\R^d$ ($d\geq3$) given a raw Dirac mixture with many components (source). By establishing a novel \vMF-based hyperspherical localized cumulative distribution (HLCD), a manifold-adaptive hyperspherical \CvM distance (HCvMD) is proposed to measure the statistical divergence between two Dirac mixtures on hyperspheres. The underlying distribution is reapproximated by a target Dirac mixture from the source in the sense of least HCvMD via Riemannian optimization. Built upon the proposed HDMR, a reapproximation and reconstruction (\rar) procedure is developed for reobtaining the underlying densities in the continuous form represented by \vMF mixtures. We further integrate the HDMR method into a new hyperspherical reapproximation discrete filter (HRDF) for nonlinear estimation on hyperspheres with unknown and complex system noise. Numerical evaluations show that it yields superior tracking performance over filtering methods using a sequential Monte Carlo scheme or parametric model.

\section{Hyperspherical Dirac Mixture Reapproximation for Efficient Discrete Modeling} \label{sec:HDMR}

Discrete probabilistic modeling based on Dirac mixtures has the advantage of approximating arbitrary underlying distributions in a non-parametric manner. However, models built upon raw samples (\textit{source}) usually have a large degree of randomness and contain redundant information. Fitting a parametric PDF to the raw samples imposes strong assumptions. Thus, it is our goal to reapproximate the underlying source samples with another Dirac mixture with fewer components at more representative locations (\textit{target}). Consequently, Dirac mixture-based discrete modeling becomes deterministic and more efficient while remaining applicable for representing arbitrary hyperspherical distributions.

\subsection{Problem Formulation} \label{subsec:dm}

Given a source set of large cardinality $\nts$ on the unit hypersphere  $\rX=\{\rux_i\}_{i=1}^\nts\subset\Sbb^{d-1}\subset\R^d$ and its associated sample weights $\{\rom_i\}_{i=1}^\nts$, a Dirac mixture distribution can be formulated with each of its components located at each sample as
\begin{equation} \label{eq:sS}
	f_\rX(\ux)=\sum_{i=1}^\nts\rom_i\,\delta(\ux-\rux_i)\,,\eqwith\sum_{i=1}^\nts{\rom_i}=1\,.
\end{equation}
Here, $\delta(\ux-\ux_i)$ denotes the Dirac delta function specifying the mass cluster around each sample, and the sample weights are normalized. A target set $\sX=\{\ux_i\}_{i=1}^n\subset\Sbb^{d-1}$ with a much smaller cardinality $n\ll\nts$ is desired to reapproximate the underlying uncertainty in the form of the following Dirac mixture
\begin{equation} \label{eq:tS}
	f_{\sX}(\ux)=\sum_{i=1}^{n}\omega_i\,\delta(\ux-\ux_i)\,,\eqwith\sum_{i=1}^n{\omega_i}=1\,.
\end{equation}
Intuitively, the reapproximation scheme can be established upon minimizing a certain statistical divergence between the source and target sets shown as follows
\begin{equation} \label{eq:objective}	
\sX^*=\operatorname*{arg\,min}_{\sX\subset\Sbb^{d-1}}{\mD\big(\rX,\sX}\big)\,.
\end{equation}
Designing a proper distance measure $\mD(\rX,\sX)$, however, is non-trivial\footnote{\neqref{eq:objective} also takes sample weights into account, though they are not written as function inputs for conciseness.}. Since the two Dirac mixtures have no joint support, directly measuring the statistical distance between the two discrete densities, e.g., by means of Kullback--Leibler divergence or Hellinger distance, is infeasible. Measuring their divergence via an empirical distribution function is applicable. However, since the conventional cumulative distribution is not unique in multi-dimensional spaces, modifications are required. In~\cite{AT15_Hanebeck}, the so-called localized cumulative distribution (LCD) was proposed for multivariate random quantities in Euclidean spaces by integrating an isotropic Gaussian kernel function. It then yields a unique and symmetric characterization of the difference between two discrete models over continuous domains via generalizing the \CvM distance. The remainder of this section will be focused on establishing this basic workflow to the hyperspherical domain.

\subsection{Hyperspherical Localized Cumulative Distribution Using \vMF-Based Kernel} \label{subsec:lcd}
Considering the geometric structure of $\Sbb^{d-1}$, we define the hyperspherical localized cumulative distribution (HLCD) with reference to the \vMF distribution~\cite{mardia2009directional} shown below.
\begin{Definition} \label{def:hlcd}
	Suppose a hyperspherical random variable $\urx\in\Sbb^{d-1}$ modeled by a probability density function $f:\Sbb^{d-1}\rightarrow\R^+$. Its corresponding hyperspherical localized cumulative distribution (HLCD) is given in the form of
	\begin{equation} \label{eq:hlcd}
	\mF(\ual,\tau)=\int_{\Sbb^{d-1}}f(\ux)\,\kappa(\ux;\ual,\tau)\dd\ux\,.
	\end{equation}
	$\kappa(\ux;\ual,\tau)=\exp(\tau\ual^\top\ux)$ denotes the \vMF-like kernel with $\ual\in\Sbb^{d-1}$ and $\tau>0$ being its location and concentration, respectively.
\end{Definition}

The \vMF kernel has an isotropic dispersion and quantifies uncertainty on unit hyperspheres w.r.t.\  the geodesic curve length\footnote{Here, the term geodesic refers to local properties on the manifold.}. The HLCD of the source Dirac mixture can then be derived as
\begin{equation} \label{eq:sLCD}
\mathring{\mF}(\ual,\tau)=\int_{\Sbb^{d-1}}\sum_{i=1}^\nts\rom_i\,\delta(\ux-\rux_i)\,\kappa(\ux ;\ual,\tau)\dd\ux=\sum_{i=1}^\nts\rom_i\,\kappa\,(\rux_i;\ual,\tau)\,.
\end{equation}
Similarly, the target HLCD follows
\begin{equation}\label{eq:tLCD}
\mF(\ual,\tau)=\int_{\Sbb^{d-1}}\sum_{i=1}^n\omega_i\,\delta(\ux-\ux_i)\,\kappa(\ux ;\ual,\tau)\dd\ux=\sum_{i=1}^n\omega_i\,\kappa\,(\ux_i ;\ual,\tau)\,.
\end{equation}
The proposed \vMF-based HLCD induces a local and manifold-adaptive characterization of the underlying continuous distribution in hyperspherical domains. Based thereon, the following introduces a smooth metric to measure the statistical divergence between two hyperspherical Dirac mixtures.

\subsection{Hyperspherical \CvM Distance} \label{subsec:distance}
Essentially, the HLCD is a smooth characterization of Dirac mixtures w.r.t.\ a certain kernel location and concentration value. Given two Dirac mixture HLCDs, we modify the conventional \CvM distance by integrating their squared difference over all possible kernel locations and concentration values. The resulting hyperspherical \CvM distance (HCvMD) is shown as follows
\begin{equation} \label{eq:cvm}
	\mD(\sX,\rX)=\int_{\R_+}\mW(\tau)\int_{\Sbb^{d-1}}\Big(\mF(\ual,\tau)-\rmF(\ual,\tau)\Big)^2\,\dd\ual\dd\tau\,.
\end{equation}
$\mW(\tau)$ denotes the weighting function controlling the impact of concentration $\tau$ on the distance metric. By minimizing the statistical divergence in\neqref{eq:cvm} under the geometric constraint of the manifold, the induced target Dirac mixture reapproximates the underlying uncertainty through optimally preserving the probability mass characterized by the source. Computation of the HCvMD in\neqref{eq:cvm} then boils down to three parts, namely,
\begin{equation*}
\begin{aligned}
\mD(\sX,\rX)&=\int_{\R_+}\mW(b)\int_{\Sbb^{d-1}}\Big(\mF(\ual,\tau)-\rmF(\ual,\tau)\Big)^2\,\dd\ual\dd\tau\eqqcolon\mD_1(\sX)-2\mD_2(\sX,\rX)+\mD_3(\rX)\,,\eqwith
\end{aligned}
\end{equation*}
\begin{equation*}
\begin{aligned}
\mD_1(\sX)&=\int_{\R_+}\mW(\tau)\int_{\Sbb^{d-1}}\mF^2(\ual,\tau)\dd\ual\dd\tau\,,\\
\mD_2(\sX,\rX)&=\int_{\R_+}\mW(\tau)\int_{\Sbb^{d-1}}\mF(\ual,\tau)\,\rmF(\ual,\tau)\dd\ual\dd\tau\,,\\
\mD_3(\rX)&=\int_{\R_+}\mW(\tau)\int_{\Sbb^{d-1}}\rmF^2(\ual,\tau)\dd\ual\dd\tau\,.
\end{aligned}
\end{equation*}
By substituting the source and target LCDs in\neqref{eq:sLCD} and\neqref{eq:tLCD}, respectively, we obtain 
\begin{equation} \label{eq:Ds}
\begin{aligned}
\mD_1(\sX)&=\sum_{i=1}^n\sum_{j=1}^n\omega_i\,\omega_j\int_{\R_+}\mW(\tau)\int_{\Sbb^{d-1}}\kappa(\ux_i;\ual,\tau)\,\kappa(\ux_j;\ual,\tau)\dd\ual\dd\tau\,,\\
\mD_2(\sX,\rX)&=\sum_{i=1}^n\sum_{r=1}^\nts\omega_i\,\rom_r\int_{\R_+}\mW(\tau)\int_{\Sbb^{d-1}}\kappa(\ux_i;\ual,\tau)\,\kappa(\rux_r;\ual,\tau)\dd\ual\dd\tau\,,\\
\mD_3(\rX)&=\sum_{r=1}^\nts\sum_{s=1}^\nts\rom_r\,\rom_s\int_{\R_+}\mW(\tau)\int_{\Sbb^{d-1}}\kappa(\rux_r;\ual,\tau)\,\kappa(\rux_s;\ual,\tau)\dd\ual\dd\tau\,.
\end{aligned}
\end{equation}
Summands in\neqref{eq:Ds} take double integrals of multiplied kernels evaluated at pairwise samples. Their general expression, the so-called \textit{HCvMD unit}, is abstracted as
\begin{equation*}
\mQ(\uu,\uv)=\int_{\R_+}\mW(\tau)\,\mP(\uu,\uv,\tau)\dd\tau\,,\eqwith\mP(\uu,\uv,\tau)=\int_{\Sbb^{d-1}}\kappa(\uu;\ual,\tau)\,\kappa(\uv;\ual,\tau)\dd\ual
\end{equation*}
being the first-layer integral over the kernel locations and $\uu,\uv\in\Sbb^{d-1}$ two arbitrary samples.  Given the definition in Def.~\ref{def:hlcd}, the product of two \vMF kernels results in another \vMF kernel with a re-scaled concentration of $\tau\,\Vert\uu+\uv\Vert=\tau\,(2+2\,\uu^\top\uv)^{1/2}$. Thus, the integral over kernel locations can be reformulated into 
\begin{equation*}
\mP(\uu,\uv,\tau)=\int_{\Sbb^{d-1}}\exp\big(\tau(\uu+\uv)^\top\ual\big)\dd\ual=\int_{\Sbb^{d-1}}\kappa\Big(\alpha;\widehat{\uu+\uv},\tau\,(2+2\,\uu^\top\uv)^{1/2}\Big)\dd\ual\,,
\end{equation*}
with $\widehat{\uu+\uv}$ being the unit vector normalized from $\uu+\uv$. The integration yields the normalization constant of the \vMF distribution $\mathcal{VMF}\big(\ual;\widehat{\uu+\uv},\tau(2+2\,\uu^\top\uv)^{1/2}\big)$, i.e.,
\begin{equation}\label{eq:intP}
	\mP(\delta,\tau)=\frac{(2\pi)^{d/2}\,\mI_{d/2-1}\big(\tau\sqrt{2+2\delta}\,\big)}{\big(\tau\sqrt{2+2\delta}\,\big)^{d/2-1}}\,,\eqwith\delta=\uu^\top\uv
\end{equation}
measuring the distance between $\uu$ and $\uv$ consistently to hyperspherical geometry. $\mI_{d/2}(\cdot)$ denotes the modified Bessel function of the first kind and of order $d/2$. The second integral over the concentration $\tau$ then follows
\begin{equation}\label{eq:intQ}
\begin{aligned}
\mQ(\delta)&=\int_{\R_+}\mW(\tau)\,\mP(\delta,\tau)\dd\tau=(2\pi)^{d/2}\int_{\R_+}\mW(\tau)\frac{\,\mI_{d/2-1}\big(\tau\sqrt{2+2\delta}\,\big)}{\big(\tau\sqrt{2+2\delta}\,\big)^{d/2-1}}\dd\tau\,.
\end{aligned}
\end{equation}

\subsection{Designing the Weighting Function} \label{subsec:weight}
The weighting function $\mW(\tau)$ in\neqref{eq:cvm} can be chosen flexibly as long as its product with $\mP(\delta,\tau)$ in\neqref{eq:intQ} is integrable. However, for efficient optimization, it is preferable to design the weighting function in a way such that\neqref{eq:intQ} yields a closed-form integral. Existing CvM-based metrics deployed fixed weighting scheme and still requires approximation to guarantee the desired analytic form~\cite{AT15_Hanebeck,ACC13_Gilitschenski,IFAC20_Frisch}. To address this issue, we select a weighting function of the form
\begin{equation} \label{eq:weight}
\mW(\tau)=e^{-\epsilon\tau}\tau^{d/2-2}
\end{equation}
with $\epsilon>2$ being a domain-related parameter. The integral over kernel concentration $\tau$ in\neqref{eq:intQ} then follows
\begin{equation*}
	\mQ(\delta)=\frac{(2\pi)^{d/2}}{(\sqrt{2+2\delta})^{d/2-1}}\int_0^\infty e^{-\epsilon\tau}\tau^{-1}\mI_{d/2-1}\Big(\tau\sqrt{2+2\delta}\Big)\dd\tau\,.
\end{equation*}
As given in~\cite[17.13.112]{gradshteyn2014table}, for functions of form $t^{-1}\mI_{\nu}(at)$, their Laplace transforms take the following expression
\begin{equation*}
\mathscr{L}\big\{t^{-1}\mI_{\nu}(at)\big\}(\epsilon)=\int_0^\infty e^{-\epsilon t}t^{-1}\mI_{\nu}(at)\dd t=\nu^{-1}a^\nu\Big(\sqrt{\epsilon^2-a^2}+\epsilon\Big)^{-\nu}\,,
\end{equation*}
under the condition $\mathrm{Re}(\epsilon)>\mathrm{Re}(a)$. Give that $a=\sqrt{2+2\delta}$ with $\delta\in[\,-1,1\,]$, this condition is satisfied and we obtain the HCvMD unit\neqref{eq:intQ} in closed form
\begin{equation}\label{eq:intQclose}
\mQ(\delta)=\frac{(2\pi)^{d/2}}{(\sqrt{2+2\delta}\,)^{d/2-1}}\mathscr{L}\Big\{\tau^{-1}\mI_{d/2-1}\Big(\tau\sqrt{2+2\delta}\Big)\Big\}(\epsilon)=\frac{(2\pi)^{d/2}}{d/2-1}\Big(\sqrt{\epsilon^2-2\delta-2}+\epsilon\Big)^{1-d/2}\,.
\end{equation}
Consequently, HCvMD components between the source and target Dirac mixtures in\neqref{eq:Ds} are obtained as follows 
\begin{equation}\label{eq:DsClose}
\begin{aligned}
\mD_1(\sX)&=\sum_{i=1}^n\sum_{j=1}^n\omega_i\,\omega_j\,\mQ(\ux_i,\ux_j)=\frac{(2\pi)^{d/2}}{d/2-1}\sum_{i=1}^n\sum_{j=1}^n\omega_i\,\omega_j\Big(\sqrt{\epsilon^2-2\,\ux_i^\top\ux_j-2}+\epsilon\Big)^{1-d/2}\,,\\
\mD_2(\sX,\rX)&=\sum_{i=1}^n\sum_{r=1}^\nts\omega_i\,\rom_r\,\mQ(\ux_i,\rux_r)=\frac{(2\pi)^{d/2}}{d/2-1}\sum_{i=1}^n\sum_{r=1}^\nts\omega_i\,\rom_r\Big(\sqrt{\epsilon^2-2\,\ux_i^\top\rux_r-2}+\epsilon\Big)^{1-d/2}\,,\\
\mD_3(\rX)&=\sum_{r=1}^\nts\sum_{s=1}^\nts\rom_r\,\rom_s\,\mQ(\rux_r,\rux_s)=\frac{(2\pi)^{d/2}}{d/2-1}\sum_{r=1}^\nts\sum_{s=1}^\nts\rom_r\,\rom_s\Big(\sqrt{\epsilon^2-2\,\rux_r^\top\rux_s-2}+\epsilon\Big)^{1-d/2}\,.
\end{aligned}
\end{equation}
The selected weighting function in\neqref{eq:weight} enables analytical and theoretically sound solution of computing HCvMD for arbitrary dimensions of unit hyperspheres ($d\geq3$ for $\Sbb^{d-1}$). The measured statistical divergence only depends on the relative distance between all possible Dirac components (via the inner product) and not their individual absolute locations.  In this sense, it provides a symmetric and unique measure to quantify the difference of probability mass characterized by two hyperspherical discrete models. Moreover, it is continuous w.r.t.\ the sample locations given the resulting closed form, thereby benefiting reapproximation via the optimization procedure shown as follows.

\subsection{Reapproximation via Riemannian Optimization on Oblique Manifolds} \label{subsec:opt}
Given a hyperspherical Dirac mixture approximating its underlying distribution, we perform reapproximation by specifying\neqref{eq:objective} in the sense of the proposed HCvMD.
For implementation, we concatenate elements of $\sX$ and $\rX$ column-wise into matrices $\fX$ and $\mathring{\fX}$, respectively, and the reapproximation optimization problem is rewritten as
\begin{equation} \label{eq:opt}	
\fX^*=\operatorname*{arg\,min}_{\fX\in\OB(d,n)}{\mD(\fX,\mathring{\fX})}\,,
\end{equation}
with $\OB(d,n)\subset\R^{d\times{n}}$ being an oblique manifold~\cite{absil2009optimization}. The matrix $\fX$ stays on this manifold as the desired $n$ target samples are confined to the unit hypersphere $\Sbb^{d-1}$.

Such constrained optimization problems can be solved by conventional approaches with implicit handling of the manifold constraint, e.g., using Lagrange multipliers. An alternative for better convergence properties is the Riemannian optimization scheme~\cite{absil2009optimization}, where manifold constraints are explicitly handled without reformulation of the objective. Within this scope, approaches have been systematically established recently with successful deployment to various control applications~\cite{jiang2021freq,sato2017rie}.

\subsection{Implementation} \label{subsec:implement}
Typically, the source Dirac mixture may have either uniformly or nonuniformly weighted components, whereas components of the target are desired to be equally weighted for fast reapproximation and further use (e.g., for filtering samples should contribute equally to estimates). Because the objective in\neqref{eq:opt} is smooth and real-valued on the oblique manifold $\OB(d,n)$, we apply the Riemannian trust-region (RTR) method from Manopt~\cite{manopt} for reapproximation in the sense of least HCvMD to the source Dirac mixture. 

Deploying RTR on $\OB(d,n)$ requires both the gradients and the Hessians of the objective in the ambient space $\R^{d\times{n}}$ to attain their Riemannian counterparts on the manifold during optimization. While applying numerical approximation or automatic differentiation is possible, symbolic derivatives are (almost) always preferable for better optimization stability and efficiency. Given the closed-form expressions in\neqref{eq:DsClose}, we provide the gradients and Hessians of HCvMD in Appendix~\ref{subsec:grad}.

One remaining issue is the degree of freedom of determining the parameter $\epsilon$ in the weighting function\neqref{eq:weight}. As discussed in Sec.~\ref{subsec:weight} , any $\epsilon>2$ is theoretically valid to guarantee the resulting closed form of the HCvMD unit. In practice, however, we parameterize it according to the following continuous piecewise function for numerical stability
\begin{equation}\label{eq:epsilon}
	\epsilon= 
	\begin{cases}
	2+n^{-d}, & \text{if}\ n \leq n_\epsilon \\
	2+(d\cdot n)^{-1}, & \text{otherwise}
	\end{cases}\,.
\end{equation}
The threshold value $n_\epsilon$ is precomputed given the dimensionality $d$ and the desired target sample size $n$ w.r.t.\ function continuity. To illustrate the efficacy of the proposed hyperspherical Dirac mixture reapproximation (HDMR) method, several examples are provided.

\begin{figure}[t!]
	\centering
	\begin{tabular}{ccc}
		\includegraphics[width=0.23\textwidth]{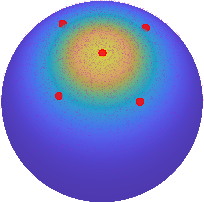} &			
		\includegraphics[width=0.23\textwidth]{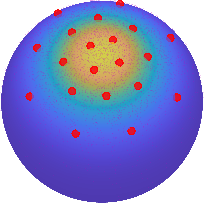} &
		\includegraphics[width=0.23\textwidth]{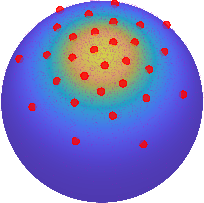} \\
		$n=5$ &  $n=20$ & $n=30$
	\end{tabular}
	\caption*{(A) $\lambda=5$}
	\begin{tabular}{ccc}
		\includegraphics[width=0.23\textwidth]{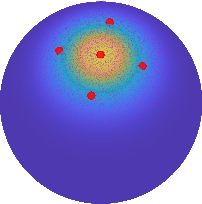} &			
		\includegraphics[width=0.23\textwidth]{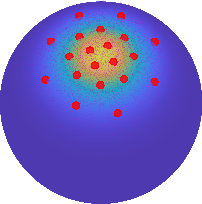} &
		\includegraphics[width=0.23\textwidth]{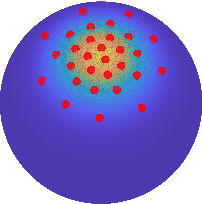} \\
		$n=5$ &  $n=20$ & $n=30$
	\end{tabular}
	\caption*{(B) $\lambda=10$}
	\begin{tabular}{ccc}
		\includegraphics[width=0.23\textwidth]{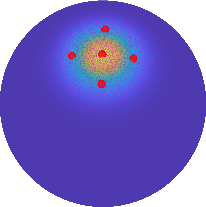} &			
		\includegraphics[width=0.23\textwidth]{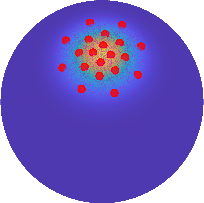} &
		\includegraphics[width=0.23\textwidth]{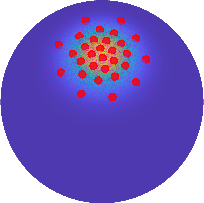} \\
		$n=5$ &  $n=20$ & $n=30$
	\end{tabular}
	\caption*{(C) $\lambda=20$}
	\caption{Reapproximations of \vMF distributions from random samples in Example~\ref{eg:uniMode}.1\,.}
	\label{fig:vmf}
\end{figure}
\begin{figure}[t!]
	\centering
	\begin{tabular}{ccc}
		\includegraphics[width=0.23\textwidth]{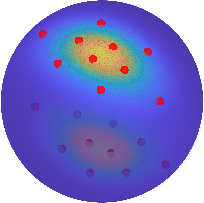} &			
		\includegraphics[width=0.23\textwidth]{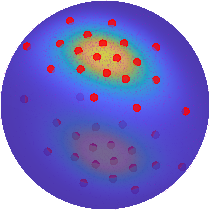} &
		\includegraphics[width=0.23\textwidth]{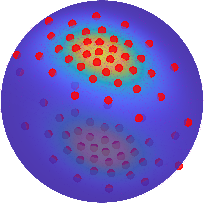} \\
		$n=20$ &  $n=40$ & $n=80$
	\end{tabular}
	\caption*{(A) $\fC=-\diag(10,4,0)$}
	\begin{tabular}{ccc}
		\includegraphics[width=0.23\textwidth]{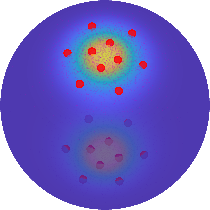} &			
		\includegraphics[width=0.23\textwidth]{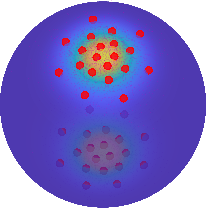} &
		\includegraphics[width=0.23\textwidth]{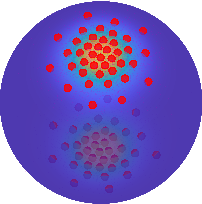} \\
		$n=20$ &  $n=40$ & $n=80$
	\end{tabular}
	\caption*{(B) $\fC=-\diag(10,10,0)$}
	\begin{tabular}{ccc}
		\includegraphics[width=0.23\textwidth]{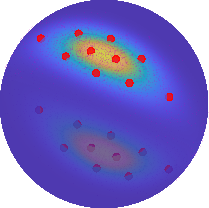} &			
		\includegraphics[width=0.23\textwidth]{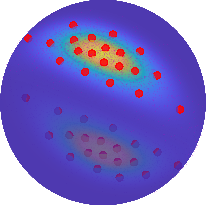} &
		\includegraphics[width=0.23\textwidth]{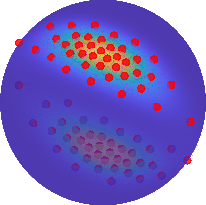} \\
		$n=20$ &  $n=40$ & $n=80$
	\end{tabular}
	\caption*{(C) $\fC=-\diag(20,4,0)$}
	\caption{Reapproximations of Bingham distributions from random samples in Example~\ref{eg:uniMode}.2\,.} 
	\label{fig:bing}
\end{figure}
\begin{figure}[t!]
	\centering
	\begin{tabular}{ccc}
		\includegraphics[width=0.23\textwidth]{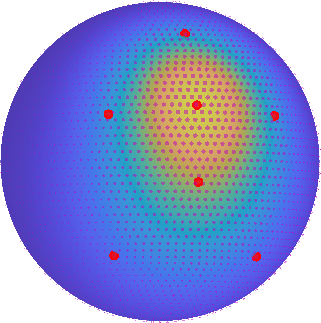} &			
		\includegraphics[width=0.23\textwidth]{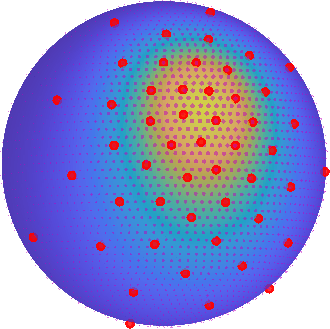} &
		\includegraphics[width=0.23\textwidth]{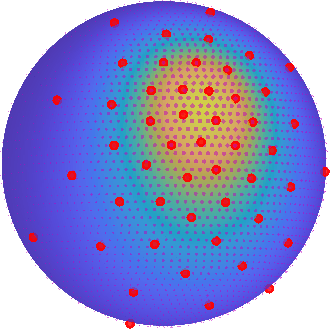} \\
		$n=7$ & $n=50$ & $n=80$
	\end{tabular}
	\caption*{(A) $\lambda_1=8,\,\lambda_2=5,\,\lambda_3=3$}
	\begin{tabular}{ccc}
		\includegraphics[width=0.23\textwidth]{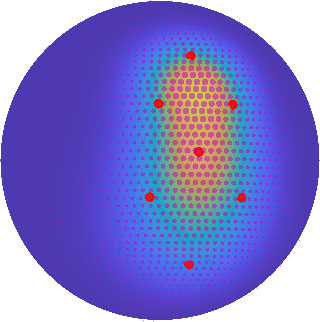} &			
		\includegraphics[width=0.23\textwidth]{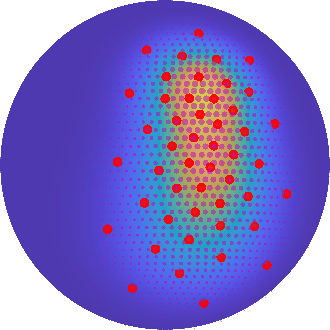} &
		\includegraphics[width=0.23\textwidth]{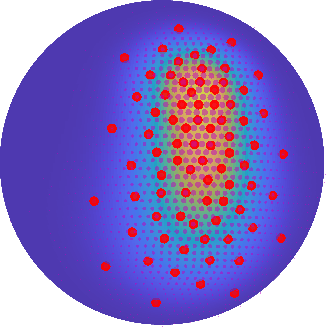} \\
		$n=7$ & $n=50$ & $n=80$
	\end{tabular}
	\caption*{(B) $\lambda_1=20,\,\lambda_2=15,\,\lambda_3=10$}
	\begin{tabular}{ccc}
		\includegraphics[width=0.23\textwidth]{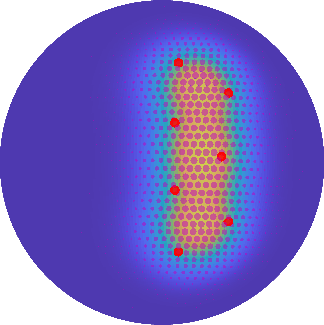} &			
		\includegraphics[width=0.23\textwidth]{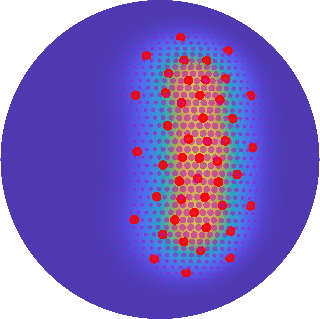} &
		\includegraphics[width=0.23\textwidth]{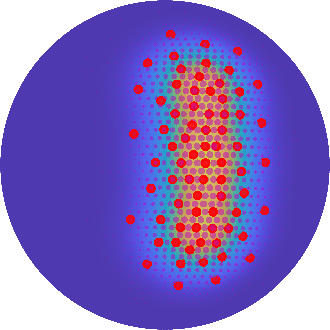} \\
		$n=7$ & $n=50$ & $n=80$
	\end{tabular}
	\caption*{(C) $\lambda_1=25,\,\lambda_2=25,\,\lambda_3=25$}
	\caption{Reapproximating distributions of complex shapes from grids on $\Sbb^2$ in Example~\ref{eg:mulMode}\,.}
	\label{fig:sphMix}
\end{figure}

\begin{Example} \label{eg:uniMode}
    Recently, parametric distributions from directional statistics have become popular statistical tools for control applications such as visual servoing~\cite{mardia2009directional,markovic2014vmf}.
	We select the following distributions as the underlying distribution for showcasing reapproximations on $\Sbb^2\subset\R^3$:
	\begin{enumerate}
		\item \vMF distributions $f_\text{vMF}(\ux;\umu,\lambda)=1/N(\lambda)\cdot\exp\,(\lambda\,\ux^\top\umu)$ with $\lambda=\{5,10,20\}$, and
		\item Bingham distributions $f_\mB(\ux;\fC)=1/N(\fC)\cdot\exp(\ux^\top\fC\,\ux)$ with parameter matrices~\cite{ECC19_Li} $\fC=\{-\mathrm{diag}(10,4,0),\,-\mathrm{diag}(10,10,0),\,-\mathrm{diag}(20,4,0)\}$.
	\end{enumerate}
	Shown in Fig.~\ref{fig:vmf}--\ref{fig:bing}, Dirac mixtures supported by $\nts=20000$ random samples (purple dots) are drawn from the underlying distributions above (therefore uniformly weighted) for reapproximation to the target Dirac mixtures of reduced components (red dots). We consider target sample sets with sizes $n=\{5,20,30\}$ and $n=\{20,40,80\}$ for the \vMF and Bingham distributions, respectively. The proposed HDMR preserves dispersion geometries of the ground truths (also well approximates the antipodal symmetry for Bingham distributions). Supports of the target mixture have deterministic locations, thereby enabling more efficient discrete probabilistic modeling compared with ones using random samples.
\end{Example}
\begin{Example}\label{eg:mulMode}
	We further synthesize distributions on $\Sbb^2\subset\R^3$ of more complex shapes with \vMF mixtures of $f_\text{vMFM}(\ux)=\sum_{i=1}^{3}1/3\cdot f_\text{vMF}(\ux;\ual_i,\lambda_i)$. We set the mode of each component to $\ual_1=1/\sqrt{5}\cdot[\,0,2,1\,]^\top$, $\ual_2=[\,0,1,0\,]^\top$ and $\ual_3=1/\sqrt{5}\cdot[\,0,2,-1\,]^\top$.
	Three sets of concentration parameters are tested, namely, $(\lambda_1,\lambda_2,\lambda_3)\in\{(8,5,3),\,(20,15,10),\,(25,25,25)\}$. Unlike the cases in Example~\ref{eg:uniMode}\,, we approximate the distribution with $\nts=5000$ Dirac components at equidistant grid points weighted by the underlying density values (with normalization)~\cite{leopardi2006partition}. For reapproximation, the target Dirac mixtures have $n=\{7,50,80\}$ components. 
	
	Shown in Fig.~\ref{fig:sphMix}, the source Dirac mixture components are plotted as purple dots with sizes proportional to their weights. Reapproximations using the proposed method yield Dirac mixtures of uniformly weighted components, with layouts being coherent to the ground truth. As the component quantities have been drastically reduced via HDMR, the proposed scheme exhibits evident advantage over the current grid-based models for efficient discrete modeling of arbitrary hyperspherical distributions.
\end{Example}

\begin{figure}
	\centering
	\begin{tabular}{ccc}
		\adjustbox{trim={.1\width} {.1\height} {.03\width} {.1\height},clip}{\includegraphics[width=0.3\textwidth]{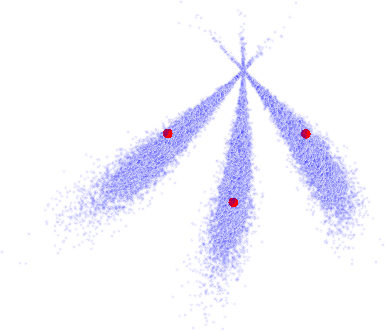}} &			
		\adjustbox{trim={.1\width} {.1\height} {0.03\width} {.1\height},clip}{\includegraphics[width=0.3\textwidth]{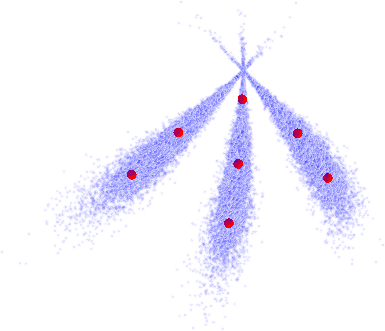}} &
		\adjustbox{trim={.1\width} {.1\height} {0.03\width} {.1\height},clip}{\includegraphics[width=0.3\textwidth]{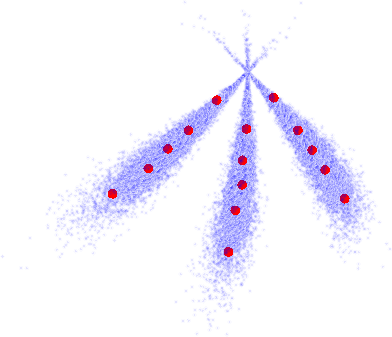}} \\
		$n=3$ & $n=7$ & $n=15$
	\end{tabular}
	\caption{Reapproximation of a quaternion distribution on $\Sbb^3$ in Example~\ref{eg:quat}\,.}
	\label{fig:quaMix}
\end{figure}

\begin{Example} \label{eg:quat}
Many control and parameter estimation tasks refer to handling quaternion-valued state vectors in stochastic processes~\cite{mayhew2011quaternion,talebi2020quaternion}. In this example, we synthesize a complicated distribution modeling random unit quaternions in the form $\urx=[\,\cos(\rth/2),\,\urn^\top\sin(\rth/2)\,]\in\Sbb^3\subset\R^4$. We assume the rotation angle to be von Mises-distributed, namely, $\btheta\sim\mathcal{VM}(\mu,\kappa)$, with $\mu=\pi/10$ and $\kappa=50$ being the mode and the concentration, respectively. The rotation axis follows a \vMF mixture distribution with density function  $f_\text{vMFM}(\un)=\sum_{i=1}^{3}1/3\cdot f_\text{vMF}(\un;\ual_i,\lambda_i)$. We configure the component means and concentrations as $\ual_1=1/\sqrt{2}\cdot[\,0,1,1\,]^\top$, $\ual_2=[\,0,1,0\,]^\top$, $\ual_3=1/\sqrt{2}\cdot[\,0,1,-1\,]^\top$ and $\lambda_1=\lambda_2=\lambda_3=200$, respectively. Given Dirac mixtures of 
$\nts=20000$ components supported by random samples drawn from the synthesis, reapproximations are performed with reduced component numbers of $n=\{3,7,15\}$. 

Shown in Fig.~\ref{fig:quaMix}, we visualize a quaternion $\ux$ by plotting a point rotated from $\uv=[\,1,0,0\,]^\top$ through $\uv^\prime=\ux\otimes\uv\otimes\ux^\circ$, with $\otimes$ being the Hamilton product and $\ux^\circ$ the conjugate of $\ux$~\cite{Fusion19_Bultmann}. Note that reapproximations are still performed on the unit hypersphere $\Sbb^3$. Though the ground truth exhibits irregular shapes of dispersion, the proposed HDMR works effectively.
\end{Example}

\section{Reapproximation-Reconstruction for Continuous Probabilistic Modeling}\label{sec:rr}
Continuous representation of unknown uncertainties is useful in many scenarios such as quantifying  density values at newly observed data or obtaining transition densities for certain recursive estimators~\cite{IFAC20_Pfaff,Fusion20_Pfaff}. As follow-up of the proposed reapproximation scheme, we introduce a method for reconstructing the underlying distribution in the form of a von Mises--Fisher mixture model, yielding the so-called \textit{reapproximation-reconstruction} (\rar) for continuous probabilistic modeling on $\Sbb^{d-1}$.

\begin{figure}[t!]
	\centering
	\begin{tabular}{cccc}
		\includegraphics[width=0.22\textwidth]{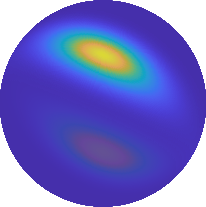} &			
		\includegraphics[width=0.22\textwidth]{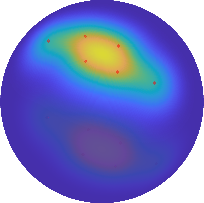} &
		\includegraphics[width=0.22\textwidth]{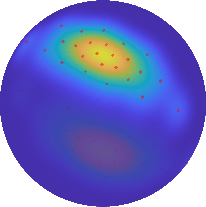} &
		\includegraphics[width=0.22\textwidth]{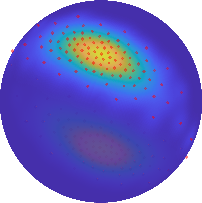} \\
		ground truth & $n=12$ & $n=50$ & $n=200$ \\
		& $\lambda^*=27.0,\mH=0.037$ & $\lambda^*=52.9,\mH=0.019$ & $\lambda^*=100.8,\mH=0.009$ 
	\end{tabular}
	 \caption*{(A) Reconstructed continuous density for Example~\ref{eg:rr}.1\,.}
	\begin{tabular}{cccc}
		\includegraphics[width=0.22\textwidth]{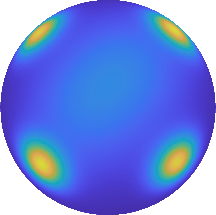} &			
		\includegraphics[width=0.22\textwidth]{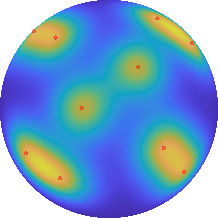} &
		\includegraphics[width=0.22\textwidth]{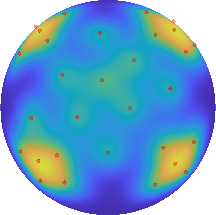} &
		\includegraphics[width=0.22\textwidth]{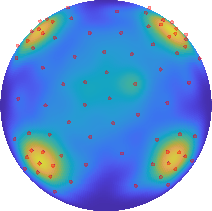} \\
		ground truth & $n=10$ & $n=30$ & $n=80$ \\
		 & $\lambda^*=19.9,\mH=0.021$ & $\lambda^*=37.5,\mH=0.013$ & $\lambda^*=56.4,\mH=0.011$ 
	\end{tabular}
 \caption*{(B) Reconstructed continuous density for Example~\ref{eg:rr}.2\,.}
	\begin{tabular}{cccc}
		\includegraphics[width=0.22\textwidth]{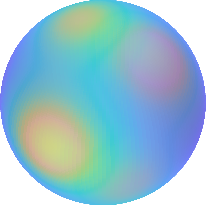} &			
		\includegraphics[width=0.22\textwidth]{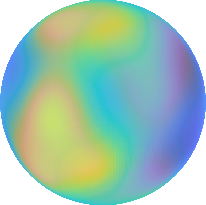} &
		\includegraphics[width=0.22\textwidth]{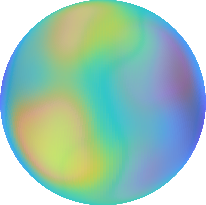} &
		\includegraphics[width=0.22\textwidth]{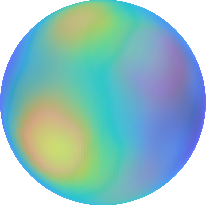} \\
	ground truth & $n=50$ & $n=100$ & $n=500$ \\
	& $\lambda^*=13.1,\mH=0.010$ & $\lambda^*=17.7,\mH=0.007$ & $\lambda^*=34.6,\mH=0.002$ 
	\end{tabular}
 \caption*{(C) Reconstructed continuous density marginalized on $\Sbb^2$ for Example~\ref{eg:rr}.3\,.}
	\caption{Continuous probabilistic modeling via proposed \rar procedure.}
	\label{fig:rr}
\end{figure}

To represent the underlying distribution in continuous form, we replace the target Dirac components from reapproximation in\neqref{eq:tS} one by one with \vMF PDFs with a single shared concentration parameter $\lambda$, i.e.,
\begin{equation}\label{eq:vmfm}
	f_\text{vMFM}(\ux;\sX,\lambda)=\sum_{i=1}^n\frac{1}{n}\,f_\text{vMF}(\ux;\ux_i,\lambda)=\sum_{i=1}^n \frac{\lambda^{d/2-1}}{n(2\pi)^{d/2}\,\mI_{d/2-1}(\lambda)}\exp(\lambda\,\ux^\top\ux_i)\,,
\end{equation}
with $\sX=\{\ux_i\}_{i=1}^n\subset\Sbb^{d-1}$. Taking the source sample set $\rX$ as observation, $\lambda$ can be obtained by means of maximum likelihood estimation (MLE) with objective formulated as log-likelihood
\begin{equation*}
\lambda^*=\operatorname*{arg\,max}_{\lambda\geq 0}\bigg\{\rom_r\ln\Big(\prod_{r=1}^\nts f_\text{vMFM}(\rux_r;\sX,\lambda)\Big)\bigg\}\\=\operatorname*{arg\,max}_{\lambda\geq 0}\bigg\{\sum_{r=1}^\nts\rom_r\ln\Big(\frac{\lambda^{d/2-1}}{\mI_{d/2-1}(\lambda)}\sum_{i=1}^n\exp(\lambda\,\rux_r^\top\ux_i)\Big)\bigg\}\,,
\end{equation*}
Here, $\rux_r\in\rX$ denotes the source sample set to be fitted. We do the substitution $\mC(\lambda)=\frac{\lambda^{d/2-1}}{\mI_{d/2-1}(\lambda)}$ and the objective function can be formulated as
\begin{equation*}
	\cl(\lambda)=\sum_{r=1}^\nts\rom_r\ln\Big(\mC(\lambda)\sum_{i=1}^n \exp(\lambda\,\rux_r^\top\ux_i)\Big)=\ln\big(\mC(\lambda)\big)+\sum_{r=1}^\nts\rom_r\ln\Big(\sum_{i=1}^n \exp(\lambda\,\rux_r^\top\ux_i)\Big)\,.
\end{equation*}
Further, its first derivative w.r.t.\ the concentration parameter $\lambda$ can be derived in the following form
\begin{equation} \label{eq:mled}
\begin{aligned}
	\cl^\prime(\lambda)&=\frac{\dd\,}{\dd\,\lambda}\bigg\{\ln\big(\mC(\lambda)\big)+\sum_{r=1}^\nts\rom_r\ln\Big(\sum_{i=1}^n \exp(\lambda\,\rux_r^\top\ux_i)\Big)\bigg\}\\
	&=\frac{\mC^\prime(\lambda)}{ \mC(\lambda)}+\sum_{r=1}^\nts\bigg(\rom_r\,\frac{\sum_{i=1}^n\rux_r^\top\ux_i\exp(\lambda\,\rux_r^\top\ux_i)}{\sum_{i=1}^n\exp(\lambda\,\rux_r^\top\ux_i)}\bigg)\,,\eqwith \mC(\lambda)=\frac{\lambda^{d/2-1}}{\mI_{d/2-1}(\lambda)}\,.
\end{aligned}
\end{equation}
The modified Bessel function of the first kind in substitute $\mC(\lambda)$ above has a well-known recurrence relation for its derivative, i.e., $\mI_{d/2-1}^\prime(\lambda)=\mI_{d/2}(\lambda)+\lambda^{-1}(d/2-1)\mI_{d/2-1}(\lambda)$~\cite{banerjee2005clustering}. Then, we obtain
\begin{equation*}
\frac{\mC^\prime(\lambda)}{\mC(\lambda)}=\frac{\mI_{d/2-1}(\lambda)}{\lambda^{d/2-1}}\cdot\frac{(d/2-1)\lambda^{d/2-2}\mI_{d/2-1}(\lambda)-\lambda^{d/2-1}\mI_{d/2-1}(\lambda)^\prime}{\big(\mI_{d/2-1}(\lambda)\big)^2}=-\frac{\mI_{d/2}(\lambda)}{\mI_{d/2-1}(\lambda)}=-\mA_d(\lambda)\,,
\end{equation*}
with $\mA_d(\lambda)$ being the Bessel function ratio that also appears in computing the mean resultant vector of the \vMF distribution~\cite{mardia2009directional}. Thus, the desired estimate of the concentration parameter can be obtained by solving equation 
\begin{equation}\label{eq:solveLambda}
	\cl^\prime(\lambda)=\sum_{r=1}^{\nts}\rom_r\,\frac{\sum_{i=1}^n\ux_i^\top\rux_r\exp(\lambda\,\ux_i^\top\rux_r)}{\sum_{i=1}^n \exp(\lambda\,\ux_i^\top\rux_r)}-\mA_d(\lambda)=0\,,
\end{equation}
As there exists no analytic solution to the equation above, we apply Newton's method with the following derivative of $\cl^\prime(\lambda)$
\begin{equation}\label{eq:mleh}
\begin{aligned}
\cl''(\lambda)&=\sum_{r=1}^\nts\rom_r\frac{\sum_{i=1}^n(\ux_i^\top\rux_r)^2\exp(\lambda\,\ux_i^\top\rux_r)}{\sum_{i=1}^n\exp(\lambda\,\ux_i^\top\rux_r)}-\sum_{r=1}^\nts\rom_r\bigg(\frac{\sum_{i=1}^n\ux_i^\top\rux_r\exp(\lambda\,\ux_i^\top\rux_r)}{\sum_{i=1}^n\exp(\lambda\,\ux_i^\top\rux_r)}\bigg)^2-\mA_d^\prime(\lambda)\,.
\end{aligned}
\end{equation}
As given in~\cite{mardia2009directional}, the first derivative of the Bessel function ratio $\mA_d(\lambda)$ in\neqref{eq:solveLambda} follows $\mA_d^\prime(\lambda)=1-\mA_d(\lambda)^2-\frac{d-1}{\lambda}\mA_d(\lambda)$. The $k_\text{th}$ iteration of the Newton's method can then be done according to
\begin{equation}
	\lambda_k=\lambda_{k-1}-\cl^\prime(\lambda_{k-1})\,\big/\,\cl''(\lambda_{k-1})\,,
\end{equation}
with $\cl^\prime(\lambda)$ and $\cl''(\lambda)$ given by\neqref{eq:mled} and\neqref{eq:mleh}, respectively. An initial guess of $\lambda$ can be obtained by averaging its upper and lower bounds, namely, $\lambda_0=(\lambda_\text{min}+\lambda_{\max})\,/\,2$, with
\begin{equation*}
\lambda_\text{min}=\mA_d^{-1}\bigg(\sum_{r=1}^{\nts}\rom_r\,\sum_{i=1}^n\ux_i^\top\rux_r\bigg)\eqand
\lambda_\text{max}=\mA_d^{-1}\bigg(\sum_{r=1}^\nts\rom_r\max_{\ux_i\in\sX}\big(\{\ux_i^\top\rux_r\}\big)\bigg)\,,
\end{equation*}
respectively. The inverse $\mA_d^{-1}$ of the Bessel function ratio $\mA_d$ can be efficiently calculated using the method in~\cite{banerjee2005clustering}. Since the variable is bounded, we also incorporate bisections to moderate the Newton steps given the upper and lower bounds above for better robustness. The following example showcases the efficacy of the full \rar scheme in various scenarios.
\begin{Example}\label{eg:rr}
We consider the following three continuous probability distributions as the underlying true density on hyperspheres.
\begin{enumerate}
	\item The Bingham distribution mentioned in Fig.~\ref{fig:bing}-(C) on $\Sbb^2$. Given an approximation of a random sampling-based Dirac mixture of $\nts=20000$ components, the \rar scheme aims to generate \vMF mixtures of $n=\{12,50,200\}$ components.
	\item A mixture of \vMF distributions  $f_\text{vMFM}(\ux)=\sum_{i=1}^{7}1/7\cdot f_\text{vMF}(\ux;\ual_i,\lambda_i)$ on $\Sbb^2$ with means $\{\ual_i\}_{i=1}^n$ and concentrations $\{\lambda_i\}_{i=1}^7$ being
	\begin{equation*}
	\bigg\{[0,0,1]^\top,\frac{1}{\sqrt{5}}[0,2,1]^\top,\frac{1}{\sqrt{5}}[0,-2,1]^\top,\frac{1}{\sqrt{5}}[2,0,1]^\top,\frac{1}{\sqrt{5}}[-2,0,1]^\top,\,\frac{1}{\sqrt{5}}[0,1,2]^\top,\frac{1}{\sqrt{5}}[0,-1,2]^\top\bigg\}
	\end{equation*}
	and $\{5,30,30,30,30,5,5\}$, respectively. Von Mises--Fisher mixtures of $n=\{10,30,80\}$ components are to be obtained via \rar given $\nts=20000$ source samples drawn randomly from the ground truth.
	\item A mixture of Bingham distributions $f_\mathcal{BM}(\ux)=\sum_{i=1}^2f_\mB(\ux;\fC_i)$ on $\Sbb^3$ with parameter matrices $\fC_1=\fC_2=-\mathrm{diag}(0,1,4,5)$. Additionally, we rotate the two Bingham components via two unit quaternions of angles $\{\theta_i\}_{i=1}^2=\{\pi/7,\pi/7\}$ and axes $\{\un_i\}_{i=1}^2=\{1/\sqrt{11}\cdot[\,3,1,1\,]^\top,1/\sqrt{11}\cdot[\,3,-1,-1\,]^\top\}$. A Dirac mixture supported by $\nts=20000$ grid points is deployed for \rar to \vMF mixtures of $n=\{50,100,500\}$ components.
\end{enumerate}
    We employ the Hellinger distance $\mH$~\cite{pardo2018statistical} to quantify the statistical divergence between the induced \vMF mixtures and the underlying true distributions. In order to visualize distributions on $\Sbb^3$ for the third case, we marginalize over the angular term of the quaternion variables as done in~\cite{birdal2020synchronizing} and plot the resulting distribution on $\Sbb^2$. As shown in Fig.~\ref{fig:rr}, the hyperspherical distributions are well-approximated in the form of \vMF mixture via the proposed \rar scheme. Further, a mixture of higher number of components exhibits larger joint concentration and better modeling fidelity (smaller $\mH$).
\end{Example}

\section{Hyperspherical Discrete Filtering via Reapproximation} \label{sec:hrdf}
Based on the proposed Dirac mixture reapproximation scheme, we further introduce the so-called hyperspherical reapproximation discrete filter (HRDF) for non-parametric recursive Bayesian estimation on hyperspheres.
\subsection{Recursive Bayesian Estimation on Hyperspheres} \label{subsec:sFiltering}
We consider the following setup of hyperspherical estimation problems. The system model is given as
\begin{equation} \label{eq:sysModel}
	\urx_{t+1}=\ua(\urx_t,\urw_t)\,,
\end{equation}
with $\urx_t,\,\urx_{t+1}\in\Sbb^{d-1}$ being the hyperspherical states and $\urw_t\in\Sbb^{d-1}$ the system noise of arbitrary form. $\ua\colon\Sbb^{d-1}\times\Sbb^{d-1}\rightarrow\Sbb^{d-1}$ denotes the transition function. The measurement model follows
\begin{equation} \label{eq:measModel}
	\urz_t=\uh(\urx_t,\urv_t)\,,
\end{equation}
with $\urz_t\in\Z,\,\urv_t\in\V$ denoting the measurement and the measurement noise, respectively. $\uh\colon\Sbb^{d-1}\times\V\rightarrow\Z$ is the observation model. Note that both the transition and the observation function are given in generic form and incorporate unit hyperspheres $\Sbb^{d-1}$ as input domains.

\subsection{Hyperspherical Reapproximation Discrete Filter} \label{subsec:filter}
Similar to existing discrete recursive estimators~\cite{ECC20_Li,TII18_Kurz}, Dirac mixture distributions are used for representing the posterior distribution discretely
\begin{equation} \label{eq:posterior}
	f_t^{\Tee}(\ux_t)=\sum_{i=1}^{n}\nu_{t,i}^\Tee\,\delta(\ux_t-\ux_{t,i}^\Tee)\,,
\end{equation}
where $\ux_{i,k}^\Tee\in\Sbb^{d-1}$ is the component location and $\nu_{t,i}^\Tee$ the corresponding weight with $\sum_{i=1}^{n}\nu_{t,i}^\Tee=1$. 

Unlike using parametric probability distributions, this allows approximating distributions of arbitrary shape. However, existing sample-based filtering approaches mostly rely on sequential Monte Carlo methods~\cite{gussy05} or equidistant grids~\cite{ECC20_Li}. The former typically requires large random sample sets to handle strong nonlinearities and high-dimensional states, while still prone to fail in certain challenging cases~\cite{LCSS21_Li}. The latter can only approximate underlying hyperspherical distributions up to a fixed resolution non-adaptively to the dispersion, and hence lacks flexibility and efficiency for modeling arbitrary uncertainties. For both variants, noise distributions $\rw_t$ should be given for system propagation. In this regard, approximations using parametric models are almost inevitable when only empirical noise data are provided. Thus, the proposed \rar can be deployed to efficient discrete modeling of unknown system noise as well as estimates for filtering in a non-parametric manner, leading to the hyperspherical reapproximation discrete filter (HRDF) elaborated as follows. 

\paragraph*{Prediction step} Given the posterior density in\neqref{eq:posterior} at step $t$, one can exploit the Chapman–Kolmogorov equation to predict the prior of step $t+1$ according to\footnote{We assume that the noise term is state-independent}
\begin{equation}\label{eq:prior}
\begin{aligned}
	f_{t+1}^\Tp(\ux_t)&=\int_{\Sbb^{d-1}}f_t^\Tee(\ux_{t})\int_{\Sbb^{d-1}}f(\ux_{t+1}\,\vert\,\uw_t,\ux_t)\,f^\uw(\uw_t)\dd\uw_t\dd\ux_t\\
	&=\sum_{i=1}^n\nu_{t,i}^\Tee\int_{\Sbb^{d-1}}f(\ux_{t+1}\,\vert\,\uw_t,\ux_{t,i}^\Tee)\,f^\uw(\uw_t)\dd\uw_t\,.
\end{aligned}
\end{equation}
Suppose the time-invariant noise distribution $f^\uw(\uw_t)$ has no symbolic form and is only approximated by numerous samples from empirical observation. We reapproximate the true distribution from this source via HDMR, which yields a noise Dirac mixture $f^\uw(\uw)=\sum_{k=1}^{n_\Tw}\nu_k^\Tw\,\delta(\uw-\breve{\uw}_k)$ of $n_\Tw$ components. Here, the component weights sum to one, i.e., $\sum_{k=1}^{n_\Tw}\nu_k^\Tw=1$. Taking the system model in\neqref{eq:sysModel},\neqref{eq:prior} turns into
\begin{equation}\label{eq:priorDis}
f_{t+1}^\Tp(\ux_{t+1})=\sum_{i=1}^n\sum_{k=1}^{n_\Tw}\nu_{t,i}^\Tee\,\nu^\Tw_{k}\, \delta\big(\ux_{t+1}-a(\ux_{t,i},\breve{\uw}_k)\big)\,,
\end{equation}
where state and noise samples are combined via Cartesian product and propagated through the system dynamics. The induced prior density function can be expressed as the following Dirac mixture of $n_\Tp=n\cdot{n_\Tw}$ components
\begin{equation}\label{eq:priorDisSimple}
f_{t+1}^\Tp(\ux_{t+1})=\sum_{r=1}^{n_\Tp}\nu_{t+1,r}^\Tp\,\delta(\ux_{t+1}-\ux_{t+1,r}^\Tp)\,.
\end{equation} 
Here, $\ux_{t+1,r}^\Tp=a(\ux_{t,i}^\Tee,\breve{\uw}_k)$ (with $r=(i-1)\cdot{n}+k$) denotes the location of each Dirac component after propagation.

\paragraph{Update step} 
Given measurement $\huz_{t+1}$ at time step $t+1$, we fuse it with the prior Dirac mixture in\neqref{eq:priorDisSimple} according to the Bayes' rule as follows
\begin{equation}\label{eq:updateDirac}
\begin{aligned}	
&f_{t+1}^\Tee(\ux_{t+1}\,\vert\,\huz_{t+1})\propto f_t^\TL(\huz_{t+1}\,\vert\,\ux_{t+1})f_{t+1}^\Tp(\ux_{t+1})=\sum_{r=1}^{n_\Tp}f_{t+1}^\TL(\huz_{t+1}\,\vert\,\ux_{t+1,s}^\Tp)\,\nu_{t+1,r}^\Tp\,\delta(\ux_{t+1}-\ux_{t+1,r}^\Tp)\,.
\end{aligned}
\end{equation}
Each prior Dirac component is reweighted by its likelihood $f_{t+1}^\TL(\huz_{t+1}\,\vert\,\ux_{t+1,s}^\Tp)$ incorporating the measurement $\huz_{t+1}$. The resulting Dirac mixture approximates the posterior distribution with $n\cdot{n_\Tp}$ nonuniformly weighted components instead of $n$ as required in\neqref{eq:posterior}. Thus, we perform reapproximation using the proposed scheme to reobtain a posterior Dirac mixture with uniform weights, with its components located geometry-adaptively to the underlying uncertainty. 

The proposed hyperspherical reapproximation discrete filter (HRDF) is summarized in Alg.~\ref{alg:HRDF} with pseudocode. In practice, we set the components of posterior and noise Dirac mixtures to be equally weighted: therefore, the presented filtering algorithm is only formulated with unweighted sample sets for brevity. Shown in Alg.~\ref{alg:HRDF}, line 1--5, the prior state sample $\sX_{t+1}^\Tp$ at step $t+1$ is obtained by propagating the previous posterior set $\sX_t^\Tee$ with noise samples $\W$ via a Cartesian product through the system dynamics. Then, each prior sample from the enlarged set is reweighted by its likelihood given the current measurement $\huz_{t+1}$ (Alg.~\ref{alg:HRDF}, line 6--8). Afterward, the proposed hyperspherical Dirac mixture reapproximation is performed to reobtain equally weighted posterior samples in a geometry-adaptive layout.
\begin{algorithm}[t]
 	\caption{\strut Hyperspherical Reapproximation Discrete Filter (HRDF)} \label{alg:HRDF}
 	\KwIn{posterior state set $\sX_{t}^\Tee=\{\ux_{t,i}^\Tee\}_{i=1}^{n}$, system noise set $\W=\{\breve{\uw}_k\}_{i=1}^{n_\Tw}$ }
 	\KwOut{posterior state set $\sX_{t+1}^\Tee=\{\ux_{t+1,i}^\Tee\}_{i=1}^{n}$}
 	$\sX_{t+1}^\Tp\gets\emptyset,\Omega_{t+1}^\Tp\gets\emptyset$\,\;
	\tcc{prediction step}
 	\For{$i \gets 1$ \KwTo $n$}{
	 	\For{$k \gets 1$ \KwTo $n_\mathrm{w}$}{
			$\ux_{t+1}^\Tp\gets a\,(\ux_{t,i}^\Tee,\breve{\uw}_k)$\,\;
			$\sX_{t+1}^\Tp\gets\{\,\sX_{t+1}^\Tp,\,\ux_{t+1}^\Tp\,\}$\,\;
		}
	}
 	\tcc{update step}
 	\For{$r\gets 1$ \KwTo $n\cdot{n_\mathrm{w}}$}{
		$\nu_{t+1,r}^\Tp\gets f_{t+1}^\TL(\huz_{t+1}\,\vert\,\ux_{t+1,r}^\Tp)$\,\;
		$\Omega_{t+1}^\Tp\gets\{\,\Omega_{t+1}^\Tp,\,\nu_{t+1,r}^\Tp\,\}$\,\;
 	}
	$\sX_{t+1}^\Tee\gets\texttt{HDMR}\,(\sX_{t+1}^\Tp,\Omega_{t+1}^\Tp,n)$\,\;
 \end{algorithm}

\section{Evaluation} \label{sec:eva}
\begin{figure}[t!]
	\centering
	\includegraphics[width=0.65\textwidth]{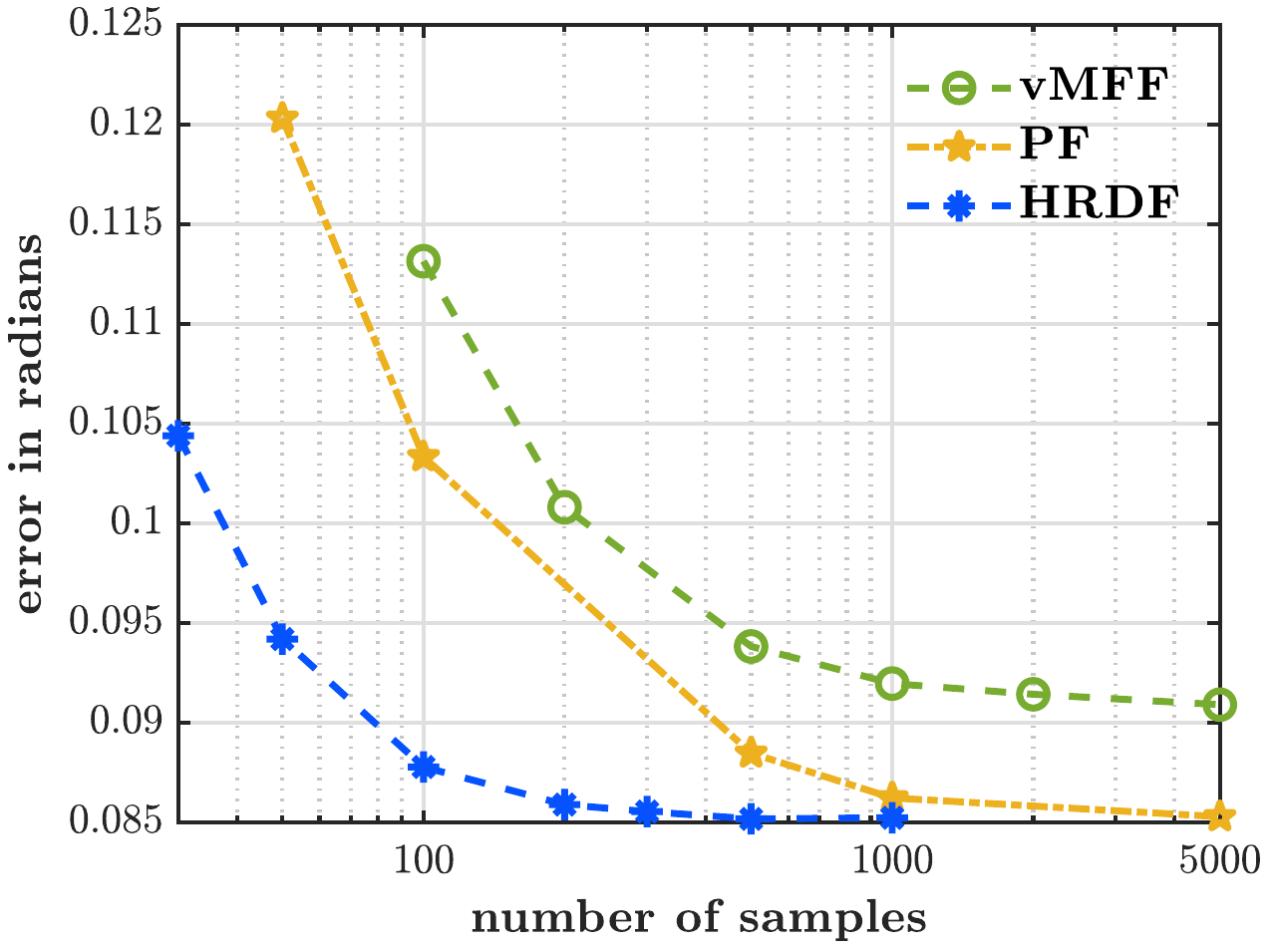}
	\caption{Error in radians over sample sizes. }
	\label{fig:error}
\end{figure}
\begin{figure}[t!]
	\centering
	\includegraphics[width=0.65\textwidth]{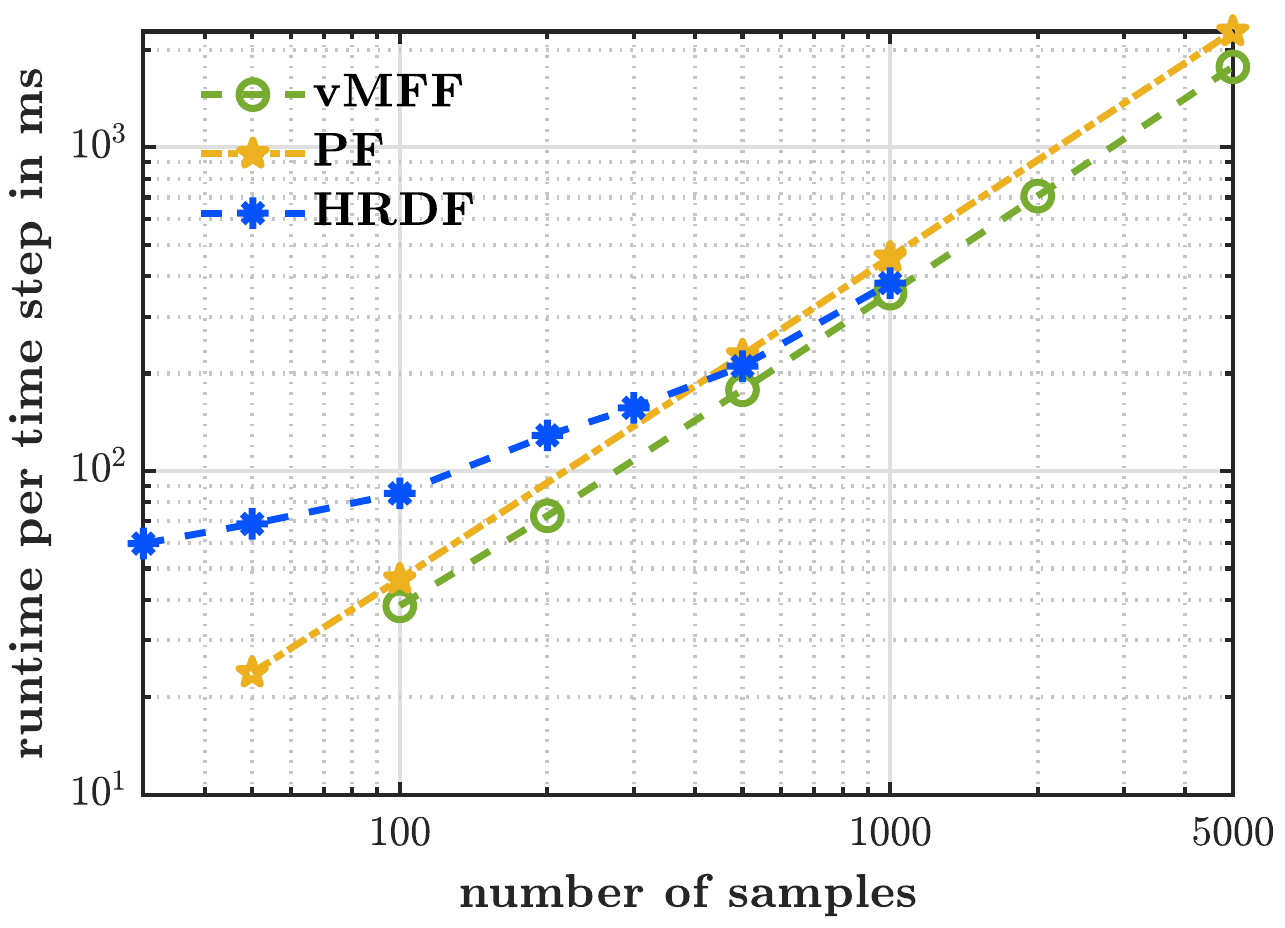} 
	\caption{Runtime in ms per time step over sample sizes.}
	\label{fig:runtime}
\end{figure}
\begin{figure}[t!]
	\centering
	\includegraphics[width=0.65\textwidth]{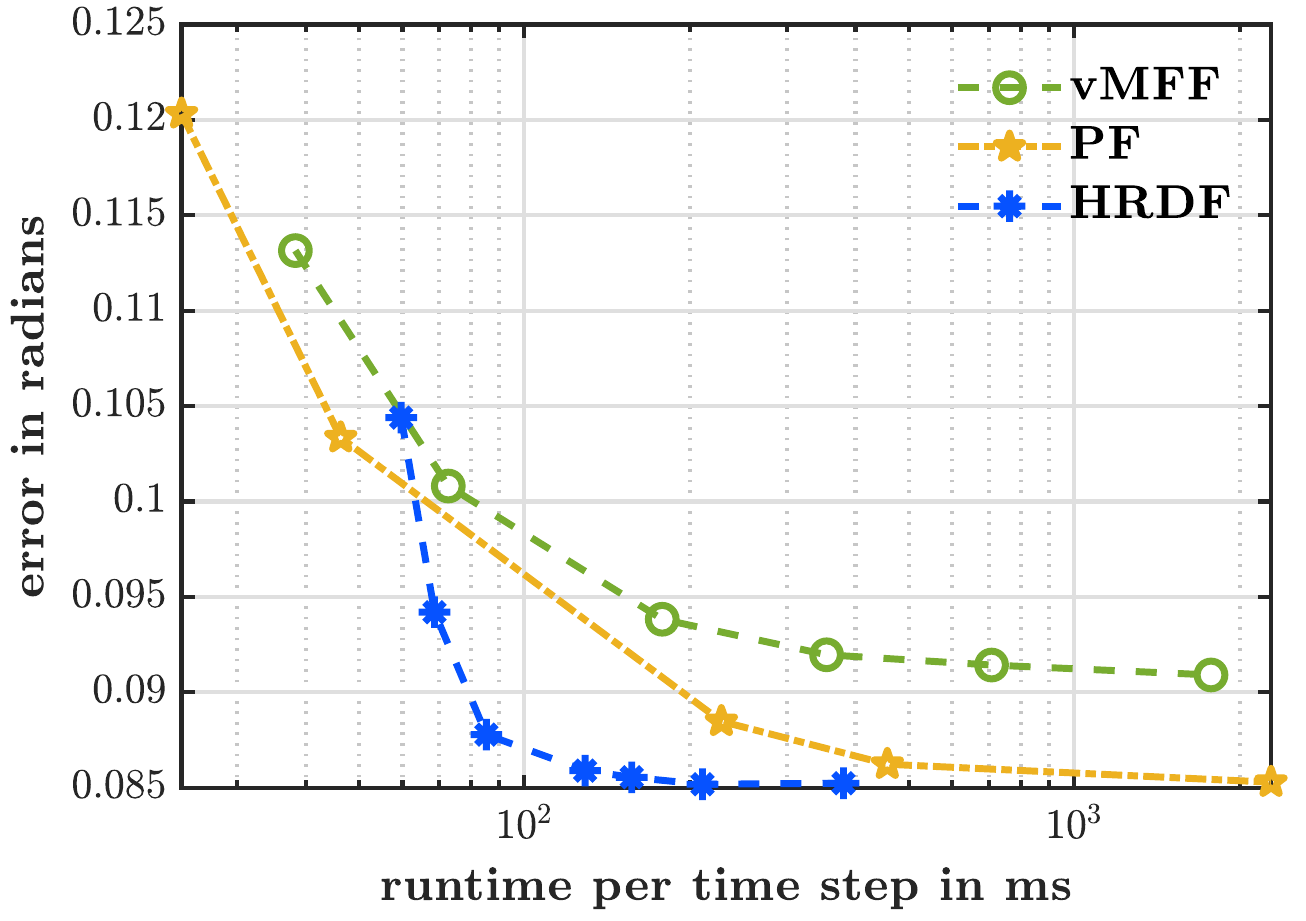} 
	\caption{Error over runtime given by the evaluated filters.}
	\label{fig:runtimeError}
\end{figure}
\begin{figure}
	\centering
	\begin{tabular}{cc}
		\includegraphics[width=0.35\textwidth]{fig/vmfm_gt.png} &
		\includegraphics[width=0.35\textwidth]{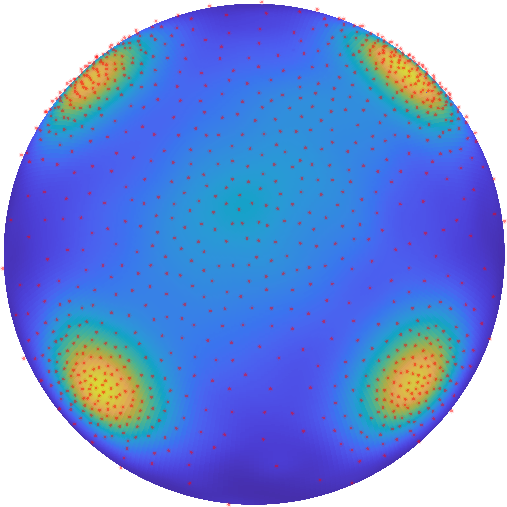} \\
		(A) true density & (B) \vMF mixture reconstruction\\
		& $\lambda=112.3$, $\mH=3.99\times10^{-3}$
	\end{tabular}
	\caption{Illustration of reapproximation fidelity for modeling the system noise in the evaluation in Sec.~\ref{sec:eva}. (A) ground truth. (B) \vMF mixture model with $n=1000$ components given by the proposed \rar approach, with the target samples plotted in red.}
	\label{fig:rr1000}
\end{figure}
We evaluate the proposed hyperspherical reapproximation discrete filter (HRDF) for hyperspherical nonlinear state estimation. Spherical motions have been commonly showcased in control  tasks~\cite{li2014unified,olfati2006warm}. Thus, we specify \neqref{eq:sysModel} to simulate motion trajectories on the unit sphere $\Sbb^2$ according to the following system model 
\begin{equation} \label{eq:systemCase}
	\urx_{t+1}=\frac{\urx_t+\urw_t}{\Vert\,\urx_t+\urw_t\,\Vert}\,,
\end{equation}
with $\urx_t,\urx_{t+1}\in\Sbb^2$ denoting the spherical states and $\rw_t\in\Sbb^2$ the system noise. It adds the state and noise variables in the Euclidean space $\R^3$ and confines their sum to the unit sphere $\Sbb^2$ via normalization. To distinguish the proposed HRDF from existing hyperspherical filters, we exploit the synthesized distribution in Example~\ref{eg:rr}.2 (plotted in Fig.~\ref{fig:rr}-(B)) as the distribution of the system noise $\rw_t$. The measurement model in\neqref{eq:measModel} is formulated into the following nonlinear form
\begin{equation}\label{eq:measCase}
\urz_t=\uh(\urx_t)+\rv_t\,,\eqwith\uh(\urx_t)=\Bigg[\arctan\bigg(\frac{\urx_{t,2}}{\urx_{t,1}}\bigg),\arctan\bigg(\frac{\urx_{t,3}}{(\urx_{t,1}^2+\urx_{t,2}^2)^{1/2}}\bigg)\Bigg]^\top
\end{equation}
observing the spherical coordinates of the state variable $\urx_t=[\,\urx_{t,1},\urx_{t,2},\urx_{t,3}\,]^\top$. The additive measurement noise $\urv_t$ is assumed to follow a zero-mean Gaussian distribution with $\urv\sim\mN(\uzero,0.01\cdot\fI_{2\times2})$, where $\fI_{2\times2}$ is a two-dimensional identity matrix.

We assume that the synthesized distribution of noise variable $\urw_t$ in\neqref{eq:systemCase} has no symbolic form and is approximated by $20000$ random samples as source. Based thereon, we exploit HDMR to obtain Dirac mixtures of $n_\Tw=\{30,50,100,200,300,500,1000\}$ components (reapproximation result for $n_\Tw=30$ visualized in Fig.~\ref{fig:rr}-(B)). Using the discrete noise distribution of different configurations, we evaluate the proposed HRDF by using a Dirac mixture comprising only $n=5$ components for representing the posterior in\neqref{eq:posterior}. To compare it with a parametric modeling-based filtering scheme, we match a \vMF distribution to the same noise source samples and deploy it to the \vMF filter (vMFF)~\cite{Sensors21_Li} using random samples of sizes ranging from $100$ to $5000$. As a baseline, a standard particle filter (PF) with sampling--importance resampling method is deployed, where the true noise density of symbolic form is exploited as the proposal distribution. The error metric for quantifying the tracking accuracy is defined as the arc length between the estimate $\hux_t$ and the ground truth $\ux_t$, i.e.,
\begin{equation*}
	\mE(\ux_t,\hux_t)=\acos(\ux_t^\top\hux_t), \eqwith \ux_t,\hux_t\in\Sbb^2\,.
\end{equation*}
We perform simulations for $5000$ Monte Carlo runs with $30$ time steps each for each run. Deviations are expressed as the root mean squared error (RMSE) of the last estimate from each run.

As shown in Fig.~\ref{fig:error}, using same numbers of samples for system propagation, the proposed HRDF delivers superior tracking accuracy over the others by maintaining only $5$ samples for discrete modeling on $\Sbb^2$. Although the PF used as baseline takes advantage of the true noise distribution, the proposed HRDF still exhibits the fastest convergence over sample sizes. It shows similar tracking accuracy using only $300$ components for noise reapproximation compared with the PF drawing $5000$ random samples from the ground truth density. This highlights the representation efficiency of geometry-adaptive Dirac mixture modeling enabled by HDMR. Moreover, there remains a clear offset between the \vMF-based filter to the baseline as it converges with larger sample sizes. This is not surprising given that the filter represents both the state estimates and the unknown noise distribution using probabilistic models of parametric form.

Fig.~\ref{fig:runtime} further illustrates the runtime of the evaluated filters over various noise sample sizes, and its relation to tracking accuracy is plotted in Fig.~\ref{fig:runtimeError}. Due to runtime overhead, e.g., allocating memory for the reapproximation, the proposed HRDF shows similar performance to the others using a small sample size. However, it quickly builds up  margins of superiority as larger noise samples are deployed. Since reapproximations of the unknown noise distribution are done offline, the proposed filter does not rely on any sampling or resampling scheme during runtime and the most time-consuming part is the reapproximation during the update step (Alg.~\ref{alg:HRDF}, line 9). This, however, can be quickly solved by the Riemannian trust-region method using the symbolic gradients and Hessians given in Sec.~\ref{subsec:implement}. Also, thanks to the reapproximation scheme, only a small number of Dirac components are needed for discrete modeling of the estimates and the noise distribution. To provide more insight in this regard, we further plot the reconstructed unknown system noise with a \vMF mixture model of $n=1000$ components given by \rar in Fig.~\ref{fig:rr1000}.  The true density is reapproximated in a geometry-adaptive manner with high fidelity, where only small statistical divergence in Hellinger distance $\mH$ exists. 

\section{Conclusions}\label{sec:conc}

In this work, we propose a novel paradigm, the hyperspherical Dirac mixture reapproximation (HDMR), for efficient discrete probabilistic modeling on unit hyperspheres of arbitrary dimensions. Taking a source Dirac mixture with many components, it reapproximates the underlying distribution with a target Dirac mixture with configurable numbers of components in a layout adaptive to the dispersion's geometry. This is performed via matching the target to the source in the sense of least hyperspherical \CvM distance (HCvMD), a new statistical divergence generalized for hyperspheres based on the proposed hyperspherical localized cumulative distribution (HLCD). Built upon the HDMR, a two-stage reapproximation and reconstruction (\rar) procedure is established for obtaining a \vMF mixture model that approximates the underlying unknown distribution in continuous form. The proposed HDMR is further integrated into the novel hyperspherical reapproximation discrete filter (HRDF) for nonlinear estimation on hyperspheres. By reapproximating hyperspherical distributions in a geometry-adaptive manner, the filter shows evident performance superiority over filters based on the Monte Carlo scheme and parametric modeling in the absence of a given noise distribution.

The presented work has considerable potential for applications in automatic control involving on-manifold random variables, where empirical data from real-world experiments can be learned by applying/modifying the proposed reapproximation scheme~\cite{Gilitschenski2020Deep}. It also provides a computational paradigm for efficient primitive extraction in cognitive control tasks such as robotic manipulation and  tactile/visual servoing~\cite{ICRA20_Li,justin2016visual,gao2020motion}. To pave the way for broader use, the reapproximation scheme needs to be tested with higher-dimensional data sets and some technical blocks, such as the weighting function in\neqref{eq:weight} and its scale-dependent parameter $\epsilon$ in\neqref{eq:epsilon}, may need modifications. The proposed filter can also be extended for more challenging tracking tasks in real-world applications, e.g., egomotion estimation based on multi-sensor fusion~\cite{RAL21_Li,hashim2019nonlinear,brossard2020Code}.

\section*{Acknowledgment}
This work is partially funded by the Helmholtz AI Cooperation Unit within the scope of
the project Ubiquitous Spatio-Temporal Learning for Future Mobility (ULearn4Mobility). The authors acknowledge support by the state of Baden-Württemberg through bwHPC.  

\bibliographystyle{IEEEtran.bst}
\bibliography{ISASPublikationen,bibliography_local}

\appendix
\section{Deriving Gradients and Hessians of HCvMD in the Ambient Space}\label{subsec:grad}
The optimization in\neqref{eq:opt} is constrained to the oblique manifold $\OB(d,n)$. For the sake of illustration, we show the symbolic forms of gradient and Hessians w.r.t.\ to each column of $\fX$ in the ambient space of $\Sbb^{d-1}$. In practice, however, both the gradients and Hessians are computed directly via matrix manipulations such that optimization on the oblique manifold can be done efficiently. 
\subsection{Pointwise Hessians in the Ambient Space of $\Sbb^{d-1}$}
We first derive the gradient of the HCvMD unit in\neqref{eq:intQclose} w.r.t.\ the sample location. For conciseness, the distance metric is reformulated as
\begin{equation*}
\begin{aligned}
\mQ(\uu,\uv)&=\frac{(2\pi)^{d/2}}{d/2-1}\bigg(\sqrt{\epsilon^2-2(1+\uu^\top\uv\big)}+\epsilon\bigg)^{1-d/2}\eqqcolon\frac{(2\pi)^{d/2}}{d/2-1}\big(\zeta(\uu,\uv)+\epsilon\big)^{1-d/2}\,,
\end{aligned}
\end{equation*}
with $\zeta(\uu,\uv)=\sqrt{\epsilon^2-2(1+\uu^\top\uv)}$.
Thus, its gradient w.r.t.\ the sample location $\uu$ follows
\begin{equation}\label{eq:gradUnit}
\frac{\partial\mQ(\uu,\uv)}{\partial\uu}=\frac{(2\pi)^{d/2}}{\zeta(\uu,\uv)\li(\zeta(\uu,\uv)+\epsilon\ri)^{d/2}}\,\uv\eqqcolon\chi(\uu,\uv)\,\uv\,, \eqwith \chi(\uu,\uv)=\frac{(2\pi)^{d/2}}{\zeta(\uu,\uv)\li(\zeta(\uu,\uv)+\epsilon\ri)^{d/2}}
\end{equation}
representing the norm of the gradient. Similarly, we obtain $\frac{\partial\mQ}{\partial\uv}=\chi(\uu,\uv)\,\uu$. 

For the HCvMD between two hyperspherical Dirac mixtures in\neqref{eq:DsClose}, its gradient w.r.t.\ each target point location $\ux_i$ takes the form $\frac{\partial\mD}{\partial\ux_i}=\frac{\partial\mD_1}{\partial\ux_i}-2\frac{\partial\mD_2}{\partial\ux_i}$, with
\begin{equation*}
\begin{aligned}
\frac{\partial\mD_1(\fX)}{\partial\ux_i}&=2\,\omega_i\sum_{j=1}^n\omega_j\frac{\partial\mQ(\ux_i,\ux_j)}{\partial\ux_i}=2\,\omega_i\sum_{j=1}^n\omega_j\,\chi(\ux_i,\ux_j)\,\ux_j\,,\\
\frac{\partial\mD_2(\fX)}{\partial\ux_i}
&=\omega_i\sum_{r=1}^\nts\rom_r\frac{\partial\mQ(\ux_i,\rux_r)}{\partial\ux_i}=\omega_i\sum_{r=1}^\nts\rom_r\,\chi(\ux_i,\rux_r)\,\rux_r\,,
\end{aligned}
\end{equation*}
being each component of the gradient, respectively. 

\subsection{Pointwise Hessians in the Ambient Space of $\Sbb^{d-1}$}
For conciseness, we first express the partial derivatives of the gradient norm $\chi(\uu,\uv)$ in\neqref{eq:gradUnit} w.r.t.\ $\uu$ and $\uv$ in the form
\begin{equation}
	\begin{aligned}
	\frac{\partial\chi(\uu,\uv)}{\partial\uu}&=\frac{(2\pi)^{d/2}\big((2+d)\,\zeta(\uu,\uv)+2\epsilon\big)}{2\,\zeta^3(\uu,\uv)(\zeta(\uu,\uv)+\epsilon)^{1+d/2}}\,\uv=\frac{(2+d)\zeta(\uu,\uv)+2\epsilon}{2(\zeta(\uu,\uv)+\epsilon)\,\zeta^2(\uu,\uv)}\chi(\uu,\uv)\,\uv\,,\\
	\frac{\partial\chi(\uu,\uv)}{\partial\uv}&=\frac{(2\pi)^{d/2}\big((2+d)\,\zeta(\uu,\uv)+2\epsilon\big)}{2\,\zeta^3(\uu,\uv)(\zeta(\uu,\uv)+\epsilon)^{1+d/2}}\,\uu=\frac{(2+d)\zeta(\uu,\uv)+2\epsilon}{2(\zeta(\uu,\uv)+\epsilon)\,\zeta^2(\uu,\uv)}\chi(\uu,\uv)\,\uu\,,
	\end{aligned}
\end{equation}
respectively. Then, the Hessian of the HCvMD unit in\neqref{eq:intQclose} can be derived by computing the partial derivatives of the gradient expression in\neqref{eq:gradUnit} w.r.t.\ $\uu$ and $\uv$ as follows
\begin{equation*}
\begin{aligned}
\frac{\partial{}^2\mQ(\uu,\uv)}{\partial\uu\,\partial\uu^\top}&=\frac{\partial\chi(\uu,\uv)}{\partial\uu}\uv^\top
=\frac{(2+d)\,\zeta(\uu,\uv)+2\epsilon}{2\,(\zeta(\uu,\uv)+\epsilon)\,\zeta^2(\uu,\uv)}\chi(\uu,\uv)\,\uv\,\uv^\top\,,\\
\frac{\partial{}^2\mQ(\uu,\uv)}{\partial\uu\,\partial\uv^\top}
&=\frac{\partial\chi(\uu,\uv)}{\partial\uv}\uv^\top+\chi(\uu,\uv)\,\fI_{d\times{d}}=\chi(\uu,\uv)\bigg(\frac{(2+d)\,\zeta(\uu,\uv)+2\epsilon}{2\,(\zeta(\uu,\uv)+\epsilon)\,\zeta^2(\uu,\uv)}\,\uu\,\uv^\top+\fI_{d\times d}\bigg).
\end{aligned}
\end{equation*}
Correspondingly, we compute the pointwise Hessians of component $\mD_1(\fX)$ in the objective w.r.t.\ the target Dirac location $\ux_i$ with the following form
\begin{equation*}
\begin{aligned}
&\frac{\partial{}^2\mD_1(\fX)}{\partial\ux_i\partial\ux_i^\top}=2\,\omega_i\sum_{j=1}^n\omega_j\frac{\partial{}^2\mQ(\ux_i,\ux_j)}{\partial\ux_i\partial\ux_i^\top}=2\,\omega_i\sum_{j=1}^n\omega_j\frac{(2+d)\,\zeta(\ux_i,\ux_j)+2\epsilon}{(\zeta(\ux_i,\ux_j)+\epsilon)\,\zeta^2(\ux_i,\ux_j)}\chi(\ux_i,\ux_j)\,\ux_j\,\ux_j^\top\,,\\
&\frac{\partial{}^2\mD_1(\fX)}{\partial\ux_i\partial\ux_j^\top}=2\,\omega_i\sum_{j=1}^n\omega_j\frac{\partial{}^2\mQ(\ux_i,\ux_j)}{\partial\ux_i\partial\ux_j^\top}=2\,\omega_i\sum_{j=1}^n\omega_j\bigg(\frac{(2+d)\,\zeta(\ux_i,\ux_j)+2\epsilon}{2\,(\zeta(\ux_i,\ux_j)+\epsilon)\,\zeta^2(\ux_i,\ux_j)}\,\ux_i\,\ux_j^\top+\fI_{d\times d}\bigg)\chi(\ux_i,\ux_j)\,.
\end{aligned}
\end{equation*}
Similarly, the pointwise Hessians of component $\mD_2(\fX)$ w.r.t.\ $\ux_i$ follows
\begin{equation*}
\begin{aligned}
\frac{\partial{}^2\mD_2}{\partial\ux_i\partial\ux_i^\top}
&=\omega_i\sum_{r=1}^\nts\rom_r\frac{\partial{}^2\mQ(\ux_i,\rux_r)}{\partial\ux_i\partial\ux_i^\top}=\omega_i\sum_{r=1}^\nts\rom_r\frac{(2+d)\,\zeta(\ux_i,\rux_r)+2\epsilon}{2\,(\zeta(\ux_i,\rux_r)+\epsilon)\,\zeta^2(\ux_i,\rux_r)}\chi(\ux_i,\rux_r)\,\rux_r\,\rux_r^\top\,.
\end{aligned}
\end{equation*}
\end{document}